\documentclass[%
 reprint,
showpacs,preprintnumbers,
 amsmath,amssymb,
 aps,
prd,
floatfix,
]{revtex4-1}

\usepackage{graphicx}
\usepackage{dcolumn}
\usepackage{bm}
\usepackage{units}
\usepackage{url}
\usepackage[mathlines]{lineno}
\usepackage{natbib}

\def \myvarProtonChi{941}
\def \myvarProtonDof{327}
\def \myvarProtonRedChi{2.9}
\def \myvarHeliumChi{3949}
\def \myvarHeliumDof{298}
\def \myvarHeliumRedChi{13.3}

\def \myvarProtonPlusRedChi{1.7}
\def \myvarProtonSolModLow{416}
\def \myvarProtonSolModHigh{555}
\def \myvarHeliumSolModLow{372}
\def \myvarHeliumSolModHigh{555}

\begin{document}


\title{Uncertainties in Atmospheric Muon-Neutrino Fluxes Arising from Cosmic-Ray Primaries}

\author{Justin Evans}
\author{Diego Garcia Gamez}
\author{Salvatore Davide Porzio}
\author{Stefan S{\"o}ldner-Rembold}
\author{Steven Wren}
\affiliation{%
School of Physics and Astronomy, University of Manchester, Oxford Road, Manchester, M13 9PL, UK 
}%

\date{\today}

\begin{abstract}
We present an updated calculation of the uncertainties on the atmospheric muon-neutrino flux arising from cosmic-ray primaries. For the first
time, we include recent measurements of the cosmic-ray primaries collected
since 2005. We apply a statistical technique that allows the determination of correlations between the parameters of the GSHL primary-flux parametrisation and the incorporation of these correlations into the uncertainty on the muon-neutrino flux. 
We obtain an uncertainty related to the primary cosmic rays of around $(5\text{--}15)\%$, depending on energy, which is about a factor of two smaller than the previously determined uncertainty. The hadron production uncertainty is added in 
quadrature to obtain the total uncertainty on the neutrino flux, which is reduced by $\approx 5\%$.
To take into account an unexpected hardening of the spectrum of primaries above energies of~\unit[100]{GeV} observed in recent measurements, we propose an alternative parametrisation and discuss its impact on the neutrino flux uncertainties.
\end{abstract}

\maketitle

\section{Introduction}
Interactions of primary cosmic rays with nuclei in the atmosphere produce neutrinos. The broad ranges covered in neutrino energy $E$ and path length $L$ allow experiments to investigate a wide range of the ratio $L/E$ and, by extension, to measure the larger neutrino mass-squared difference, $\Delta m^2_{32}$,  with great precision~\cite{ref:SuperKAtmos,ref:DeepCoreOsc}. The aim of several proposed detectors, such as PINGU~\cite{Aartsen:2014oha,TheIceCube-Gen2:2016cap}, ORCA~\cite{Adrian-Martinez:2016fdl}, and Hyper-Kamiokande~\cite{Abe:2011ts}, is the determination of the neutrino mass ordering (NMO)
by detecting atmospheric neutrinos that have traversed the Earth. To obtain a robust estimate of the significance of such an NMO determination, the uncertainties on the neutrino flux must be properly estimated.

The largest sources of uncertainty on the atmospheric neutrino flux stem from the uncertainties on the flux of cosmic-ray primaries and on the hadron production mechanisms occurring in the atmosphere. Barr \emph{et al.}~\cite{Barr:2006it} studied both effects and evaluated
the resulting uncertainty on the estimated atmospheric neutrino flux. Since 2005, new data on cosmic-ray primaries have been published that are not
included in their analysis. Here, we therefore use an updated technique to incorporate recent data and to obtain a new, more robust estimate of the uncertainties on the atmospheric muon-neutrino flux arising from 
cosmic-ray primaries. We also determine the possible improvement coming from using an alternative parametrization of the neutrino flux.

\section{Primary Spectrum}

The measured flux of primary cosmic rays covers up to 12 orders of magnitude in particle energy, ranging from \unit[$10^9$]{eV} to \unit[$10^{20}$]{eV}, which
necessitates the use of different experimental techniques for their detection~\cite{Fedynitch:2013pra}. Measurements below about \unit[$200$]{GeV} are possible with magnetic spectrometers, installed either on balloons (BESS~\cite{Shikaze:2006je,wang2002measurement,seo2000spectra,Abe:2015mga}, CAPRICE~\cite{Boezio:1998vc,Boezio:2002ha}) or spacecraft (AMS~\cite{Alcaraz:2000vp,Alcaraz:2000zz}, AMS-02~\cite{Aguilar:2015ooa,Aguilar:2015ctt}, PAMELA~\cite{Adriani:2013as,Adriani:2011cu}). Measurements at higher energies are made with balloon-borne (ATIC~\cite{Panov:2011ak}, CREAM~\cite{Yoon:2011aa}) and spacecraft mounted calorimeters (SOKOL~\cite{Ivanenko:1994gk}, PAMELA~\cite{Adriani:2013xva}). Previously, emulsion chambers have also been used (JACEE ~\cite{Christ:1998zz}, RUNJOB~\cite{Derbina:2005ta}). See, e.g., Ref.~\cite{Zatsepin:2006ci} for more discussions of the different data sets. Since these emulsion and calorimeter detectors do not capture all the energy of the primary, their energy measurements are not as accurate as that of  spectrometers~\cite{Gaisser:2002jj}. Above an energy of $10^5$~GeV the primary flux becomes extremely low, and indirect techniques must be employed using extensive ground-based air shower arrays~\cite{Fedynitch:2013pra}.

Different parametrisations of the primary cosmic-ray flux have been proposed. A description of these parametrisations can be found in Fedynitch \emph{et al.}~\cite{Fedynitch:2013pra}. Several parametrisations, like Hillas-Gaisser~\cite{Gaisser:2011cc}, Zatsepin-Sokolskaya~\cite{Zatsepin:2006ci} and poly-gonato~\cite{TerAntonyan:2000hh, hoerandel2003knee} aim at reproducing the features of the primary spectrum at higher energies, such as the \emph{knee} at $\approx$ $10^6$~GeV and the \emph{ankle} at $\approx 10^9$~GeV.

When performing a fit to cosmic-ray data including energies below $\approx 10$~GeV/n (energy per nucleon), the effect of the solar wind cannot be neglected. It causes low energy cosmic rays to be decelerated and thus changes their flux as measured on Earth, leading to a significant anti-correlation between solar activity and the low energy flux of cosmic rays~\cite{Gaisser:2000sj}. 

The GSHL parametrisation~\cite{Gaisser:2001jw,Gaisser:2002jj} is derived using low-energy data (below $10^5$~GeV/n) and is parametrised in conditions of solar minimum. A variant of the GSHL parametrisation has also been developed by Honda~\emph{et al.}~\cite{Honda:2007uxa}, where a function depending on the neutron monitor count accounts for variations of the modulation in the solar cycle.

\begin{table*}[htbp]
\centering
\begin{tabular}{c c c c | c c | c c c}
\hline\hline
$n$ & Data Set & $N_n$ & $\chi^2_n$ &
\multicolumn{2}{c|}{Acceptability} &
\multicolumn{3}{c}{Mutual Compatibility} \\
&  & & & $\Delta \chi^2_n$ & $\sqrt{2 N_n}$ &$\chi^2_{(0)\text{global}-n}$ & $\chi^2_{\text{global}-n}$ & $\Delta \chi^2_n$ \\
\hline\hline
\multicolumn{9}{c}{Proton} \\
\hline
1 & CREAMI & 10 & 57.0 & 47.0 & 4.5 & 883.8 & 876.6 & 7.2 \\
2 & AMS01 & 26 & 23.1 & 2.9 & 7.2 & 917.7 & 916.2 & 1.5 \\
3 & AMS02 & 49 & 151.2 & 102.2 & 9.9 & 789.5 & 713.9 & 75.7 \\
4 & PAMELACALO & 8 & 28.8 & 20.8 & 4.0 & 912.0 & 911.9 & 0.1 \\
5 & RUNJOB & 6 & 1.7 & 4.3 & 3.5 & 939.0 & 939.0 & 0.0 \\
6 & PAMELA & 142 & 157.3 & 15.3 & 16.9 & 783.5 & 762.6 & 20.9 \\
7 & SOKOL & 13 & 111.0 & 98.0 & 5.1 & 829.8 & 828.0 & 1.7 \\
8 & BESS & 57 & 67.2 & 10.2 & 10.7 & 873.6 & 872.0 & 1.6 \\
9 & ATIC02 & 15 & 324.7 & 309.7 & 5.5 & 616.0 & 556.9 & 59.1 \\
10 & JACEE & 5 & 18.7 & 13.7 & 3.2 & 922.1 & 921.9 & 0.2 \\
\hline
\multicolumn{9}{c}{$\chi^2_{\text{global}}=\myvarProtonChi$} \\
\hline\hline
\multicolumn{9}{c}{Helium} \\
\hline
1 & CREAMI & 10 & 38.6 & 28.6 & 4.5 & 3910.2 & 3908.7 & 1.5 \\
2 & AMS01 & 26 & 233.6 & 207.6 & 7.2 & 3715.2 & 3691.0 & 24.2 \\
3 & AMS02 & 42 & 1116.7 & 1074.7 & 9.2 & 2832.1 & 2243.6 & 588.5 \\
4 & PAMELACALO & 6 & 16.9 & 10.9 & 3.5 & 3932.0 & 3932.0 & 0.0 \\
5 & RUNJOB & 6 & 125.8 & 119.8 & 3.5 & 3823.1 & 3819.9 & 3.1 \\
6 & PAMELA & 66 & 207.3 & 141.3 & 11.5 & 3741.5 & 3679.1 & 62.4 \\
7 & SOKOL & 8 & 30.6 & 22.6 & 4.0 & 3918.2 & 3917.9 & 0.3 \\
8 & BESS & 117 & 551.0 & 434.0 & 15.3 & 3397.8 & 3367.9 & 29.9 \\
9 & ATIC02 & 15 & 1616.5 & 1601.5 & 5.5 & 2332.3 & 1217.7 & 1114.6 \\
10 & JACEE & 6 & 11.8 & 5.8 & 3.5 & 3937.0 & 3937.0 & 0.0 \\
\hline
\multicolumn{9}{c}{$\chi^2_{\text{global}}=\myvarHeliumChi$} \\
\hline\hline
\end{tabular}
\caption{Results from the acceptability and mutual compatibility methods applied to the proton and helium flux data.  The acceptability method compares the $\chi^2_{(0)n}$ for each individual experiment using the best set $S_0$ determined from the global fit. Here, $N_n$ is the number of data points for the $n$-th data set, and is also the expected value of $\chi^2$, with an uncertainty of $\pm \sqrt{2 N_n}$. We compare the uncertainty $\sqrt{2 N_n}$ with the difference between the expected ($N_n$) and calculated ($\chi^2_{(0)n}$) values, given by $\Delta \chi^2_n = \chi^2_{(0)n} - N_n$.
The mutual compatibility method compares the difference $\Delta \chi^2_n$ between $\chi^2_{(0) \text{global}-n}$, obtained by removing the $n$-th data set from the global fit and calculated by using the best set $S_0$ determined from the global fit, and
$\chi^2_{\text{global}-n}$, obtained by removing the $n$-th data set and minimising $\chi^2$ to find the best set $S_0$ after the $n$-th data set has been removed.}
\label{tab:proton}
\end{table*}

In this article, we use the original GSHL parametrisation proposed in Ref.~\cite{Gaisser:2001jw,Gaisser:2002jj}, which was also used by Barr~\emph{et al.}~\cite{Barr:2006it} to determine the uncertainties associated with the atmospheric muon-neutrino flux. It is given by
\begin{equation}
\label{eq:gshl}
\Phi(E_p) = a \left[ E_p + b \exp{ \left( c \sqrt{E_p} \right)} \right]^{-d}  \; ,
\end{equation}
where $E_p$ is the primary kinetic energy in units of GeV (GeV/nucleon for nuclear cosmic rays) and $d = \gamma + 1$ is the differential spectral index.
This parametrisation is an extension to the power-law spectrum
\begin{equation}
\Phi(E_p) = a E_p^{-d}  \; ,
\end{equation}
where $a$ is a normalization factor and $d$ the spectral index. The parameters $b$ and $c$ govern the effects of solar modulation on the primary flux and are relevant only at energies $<10$~GeV/n.

\section{Data sets and Fitting Techniques}
Barr \emph{et al.}~\cite{Barr:2006it} use data published by the AMS, BESS, CAPRICE,  JACEE, RUNJOB, and SOKOL Collaborations. 
In addition, we also include the recent cosmic-ray measurements performed by the AMS-02, ATIC, CREAM, and PAMELA Collaborations. These new data sets will change the results of the global fit of the neutrino flux and the associated uncertainties. 
We perform a global fit to the spectrum using the GSHL parametrisation (Eq.~\eqref{eq:gshl}). The uncertainties on the parameters $a$, $b$, $c$, and $d$ determined from the fit are then propagated to estimate how each parameter affects the total uncertainty on the atmospheric neutrino flux.

The time period
during which each data set is measured can be used to assign a solar modulation potential~\cite{usoskin2011solar}, which presents a series of reconstructed monthly values of modulation potential determined from neutron monitor data. The effects of the solar wind
becomes small above an energy of $\approx 10$ GeV/n~\cite{Gaisser:2000sj}. 
Below this energy we only
retain data points belonging to data sets with similar values of the solar modulation potential. The ranges are $[\myvarProtonSolModLow, \myvarProtonSolModHigh]$~MV and $[\myvarHeliumSolModLow, \myvarHeliumSolModHigh]$~MV for the proton and helium fluxes, respectively. They are chosen to include data from all different low-energy, low-solar modulation experiments (AMS, BESS, PAMELA) while keeping the inclusion range as small as possible. The ranges are also consistent with the initial assumption that the GSHL formula can be parametrised in conditions of a solar minimum. An estimate of the effects of different solar modulation values on the atmospheric neutrino flux is reported in Ref.~\cite{Gaisser:2002jj}. The difference between conditions of solar minimum and maximum amounts to an effect of $\approx 10\%$ on the atmospheric neutrino flux at Kamioka, and $\approx 20\%$ at the high latitude sites, Soudan and Sudbury.

We do not include the CAPRICE data points~\cite{Boezio:1998vc,Boezio:2002ha}, previously used in Ref.~\cite{Barr:2006it}, because of the disagreement in the mid-energy corridor ($10$--$100$~GeV/n) between the CAPRICE data and the remaining $\approx 10$ data sets in this range, which agree well with each other. The $\chi^2$/d.o.f. for the CAPRICE data set has a value of $15.8$, which is to be compared with the values for AMS ($0.88$), PAMELA ($1.10$) and BESS ($1.18$).

We use a $\chi^2$ function for the global fit,
\begin{equation}
\label{eq:globalchi}
\chi^2_{\text{global}} = \sum_n \sum_i \left( \frac{D_{n,i} - P_{n,i}}{\sigma_{n,i}} \right)^2 \; ,
\end{equation}
where $n$ labels the $18$ data sets used in the fit, $D_{n,i}$ is the value of the data point $i$ from data set $n$, with uncertainty $\sigma_{ni}$, and $P_{n,i}$ is the value expected from the GSHL parametrisation for that specific data point.

Minimizing $\chi^2_{\text{global}}$ yields the set $S_0 = \left\lbrace a_0, b_0, c_0, d_0 \right\rbrace$ of GSHL parameters that fit best all the data sets. To determine uncertainties corresponding to one standard deviation ($\sigma$), the neighbourhood of the minimum in parameter space is scanned.
For uncorrelated uncertainties, the $1\sigma$ range corresponds to
\begin{equation}
\Delta \chi^2 = \chi^2_0(a_0,b_0,c_0,d_0) - \chi^2(a,b,c,d) = 1 \;,
\end{equation}
where $\chi^2_0$ is the minimum of the $\chi^2$ function (see, e.g. Appendix A of Ref.~\cite{Stump:2001gu}).  If correlations are fully known and properly incorporated in the covariance matrix used to calculate $\chi^2$, the $\Delta \chi^2 = 1$ criterion remains valid. If correlations are not known, the value of $\Delta \chi^2$ necessary to properly estimate the uncertainties can be significantly larger than $1$. Point-to-point correlations within each data set are not known for all experiments~\cite{Alcaraz:2000vp,Alcaraz:2000zz,Aguilar:2015ooa,Aguilar:2015ctt,Panov:2011ak,Shikaze:2006je,wang2002measurement,seo2000spectra,Abe:2015mga,Boezio:1998vc,Boezio:2002ha,Yoon:2011aa,Christ:1998zz,Adriani:2013as,Adriani:2011cu,Adriani:2013xva,Derbina:2005ta,Ivanenko:1994gk}. Correlation between different data sets (experiment-to-experiment) are also unknown. The criterion $\Delta \chi^2 = 1$ would thus lead to a large underestimation of the uncertainties associated with the GSHL parameters and consequently of the uncertainties associated with the atmospheric neutrino flux. 

A similar problem arises when fitting a large number of data sets to extract parton distribution functions of the proton. We thus use an approach based on methods developed for fitting such data sets~\cite{Stump:2001gu,Pumplin:2001ct}, with
\begin{equation}
\Delta \chi^2 = \chi^2_0(a_0,b_0,c_0,d_0) - \chi^2(a,b,c,d) = T^2 \;,
\end{equation}
where $T$ is a \emph{tolerance} parameter that is determined by considerations of \emph{self-consistency}. In this article, we follow two different methods to assign a value to $T$, based on the two assumptions that the data sets used in the global fit are \emph{acceptable} and \emph{mutually compatible}. These methods take into account deviations from ideal statistical expectations observed for some data sets and indications of inconsistency between data sets if uncertainties are interpreted applying purely statistical rules~\cite{Pumplin:2001ct}.

\begin{table}[htbp]
\begin{tabular}{l c c}
\hline\hline
Parameter & Proton & Helium \\
\hline
$a$ & $14275 \pm 491$ & $531 \pm 30$ \\
$b$ & $2.44 \pm 0.10$ & $1.24 \pm 0.14$ \\
$c$ & $-0.36 \pm 0.04$ & $-0.33 \pm 0.13$ \\
$d$ ($<$ 200 GeV/n) & $2.75 \pm 0.01$ & $2.63 \pm 0.01$ \\
$\phantom{d} $ ($>$ 200 GeV/n) & $2.75 \pm 0.02$ & $2.63 \pm 0.02$ \\
\hline\hline
\end{tabular}
\caption{\label{tab:fitResults} Fit values and relative uncertainties for the four parameters of the GSHL parametrisation. Parameter values are determined with a global fit, and uncertainties are calculated using the method from Refs.~\protect\cite{Stump:2001gu,Pumplin:2001ct}. The units are chosen to give the flux $\Phi$ in
units of $[(\text{GeV}/\text{n})^{-1}\text{m}^{-2}\text{s}^{-1}\text{sr}^{-1}]$.}
\end{table}

The first criterion, {\em acceptability}, assigns a value to $T$ based on how well each individual experiment agrees with the best-fit GSHL function. We perform a global fit, yielding $\chi^2_{ \text{global}}$, to all data sets to determine the best set of parameters $S_0$. For each individual data set, $\chi^2_{(0)n}$ is then calculated. The parameters are fixed to the parameter set $S_0$. The subscript $(0)$ is used to indicate that $\chi^2$ is not minimised, but is calculated leaving the parameters fixed to the best set $S_0$ obtained from the global fit. If the $1 \sigma$ uncertainties were ideal, we would expect each $\chi^2_{(0)n}$ to be within the range $N_n \pm \sqrt{2 N_n}$, where $N_n$ is the number of data points in the data set labelled $n$. Large deviations of $\chi^2_{(0)n}$ from the expected value can be attributed to unknown systematic uncertainties or unexpected large fluctuations. To take these factors in account, we must anticipate a tolerance for $\chi^2_{(0) \text{global}}$ that is larger than that for an ``ideal" $\chi^2$ function.

The second criterion, {\em mutual compatibility}, determines a value for $T$ based on the degree of consistency between an individual data set and the remaining data sets. This is done by removing each of the data sets in turn from the analysis and calculating $\chi^2_{(0) \text{global}-n}$. The set of parameters $S_0$ is then adjusted to minimise $\chi^2$ and obtain $\chi^2_{\text{global}-n}$. The difference between the two values is calculated as
\begin{equation}
\Delta \chi^2 = \chi^2_{(0) \text{global}-n} - \chi^2_{\text{global}-n} \; .
\end{equation}
This is equivalent to asking how large $\Delta \chi^2$ should be to accommodate the return of the removed data set.

\begin{table}[htbp]
\begin{tabular}{c | c c c c}
\hline\hline
\multicolumn{5}{c}{Proton} \\
\hline
Parameter & $a$ & $b$ & $c$ & $d$ \\
\hline
$a$ & $\phantom{-}1\phantom{.00}$  & $-0.38$ & $\phantom{-}0.72$  & $\phantom{-}0.98$  \\
$b$ & & $\phantom{-}1\phantom{.00}$ & $-0.85$  & $-0.33$  \\
$c$ & & & $\phantom{-}1\phantom{.00}$ & $\phantom{-}0.67$  \\
$d$ & & & & $\phantom{-}1\phantom{.00}$  \\
\hline\hline
\multicolumn{5}{c}{Helium} \\
\hline
$a$ & $\phantom{-}1\phantom{.00}$  & $-0.56$ & $\phantom{-}0.79$  & $\phantom{-}0.98$  \\
$b$ & & $\phantom{-}1\phantom{.00}$ & $-0.91$  & $-0.50$  \\
$c$ & & & $\phantom{-}1\phantom{.00}$ & $\phantom{-}0.73$  \\
$d$ & & & & $\phantom{-}1\phantom{.00}$  \\
\hline\hline
\end{tabular}
\caption{Correlation matrices for the four parameters of the GSHL parametrisation.}
\label{tab:corrMat}
\end{table}

The tolerance $T$ is a global factor that applies to all the individual $\chi^2$ (i.e., there are no weights associated with the data sets). Finding the value for the tolerance factor $T$ allows for the determination of the uncertainties associated with the parameters $a$, $b$, $c$, and $d$. In Ref.~\cite{Barr:2006it} these uncertainties are propagated into the atmospheric neutrino flux by performing several Monte Carlo simulations of the atmospheric neutrino flux and altering each time the values of the four parameters by $1 \sigma$ to obtain the variation in neutrino flux as a function of energy. The deviations obtained for each parameter are then added in quadrature, without taking correlations between primary flux parameters into account.

A change in primary flux across the relevant regions of parameter space directly translates into an identical change in neutrino flux~\cite{Barr:2006it}. We therefore use the uncertainties as a function of energy obtained in Ref.~\cite{Barr:2006it} to determine new uncertainties on the muon atmospheric neutrino flux by scaling the uncertainty functions according to the results obtained using the method from Ref.~\cite{Stump:2001gu,Pumplin:2001ct}. 

An advantage of this method is that the parameters and their uncertainties are determined with a fit. It thus provides a more robust procedure for the determination of the uncertainties associated with the atmospheric neutrino flux. Future updates to cosmic-ray data can easily be incorporated into the analysis, thus making the response to new data more rapid. Furthermore, the covariance matrix for the four parameters can be extracted from the fit, thus allowing the determination of the correlations between the parameters $a$, $b$, $c$, and $d$. The results, discussed in Sec.~\ref{sec:results}, show that the parameters are highly correlated, and a proper treatment of the correlations is thus necessary. We provide the code used for fitting the cosmic-ray data at Ref.~\footnote{{\tt https://sdporzio.github.io/CRFitter/}}.

\section{Results}
\label{sec:results}

\begin{figure*}[htbp]
(a) 
\includegraphics[width=0.465\textwidth]{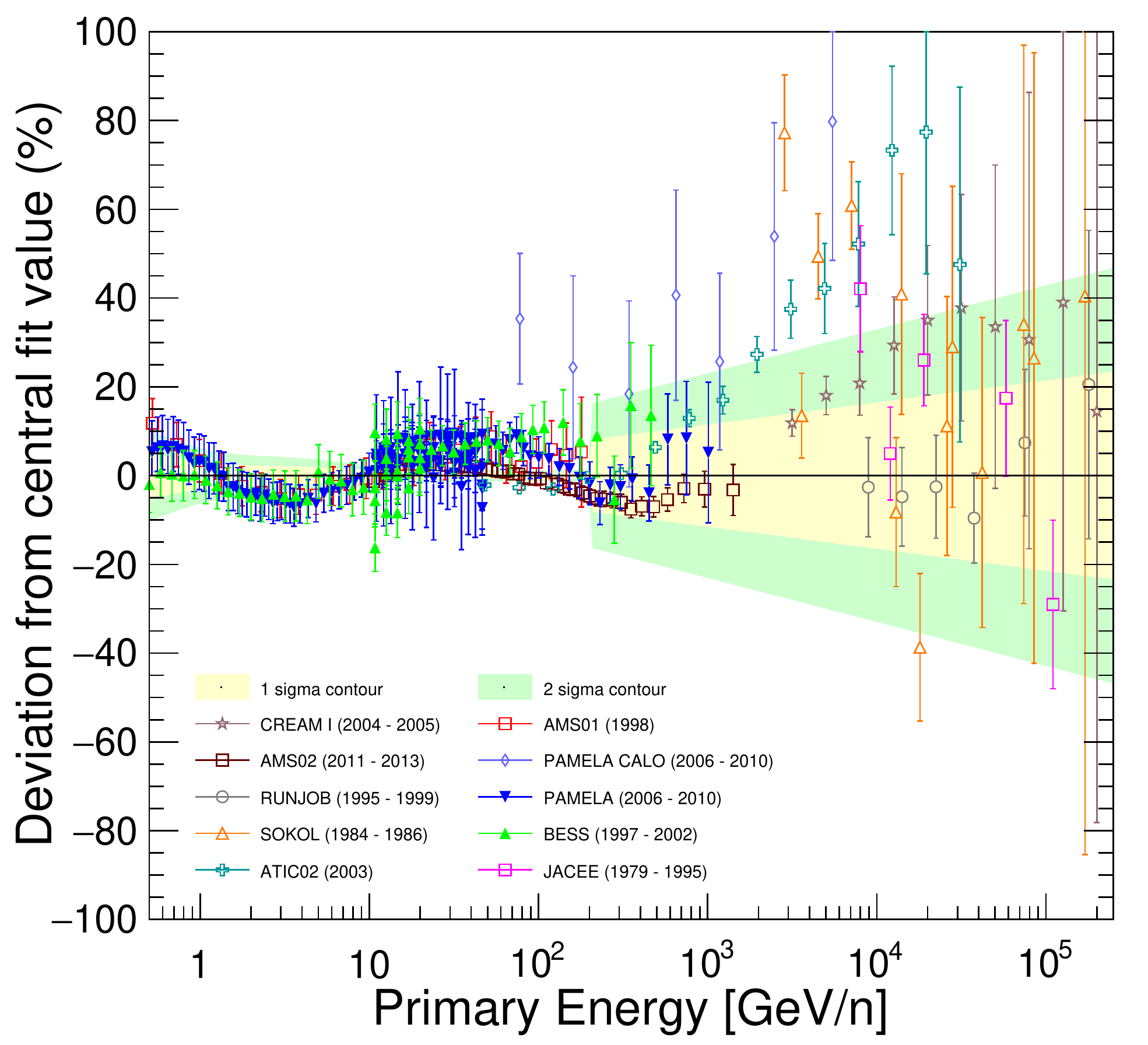}
(b)
\includegraphics[width=0.465\textwidth]{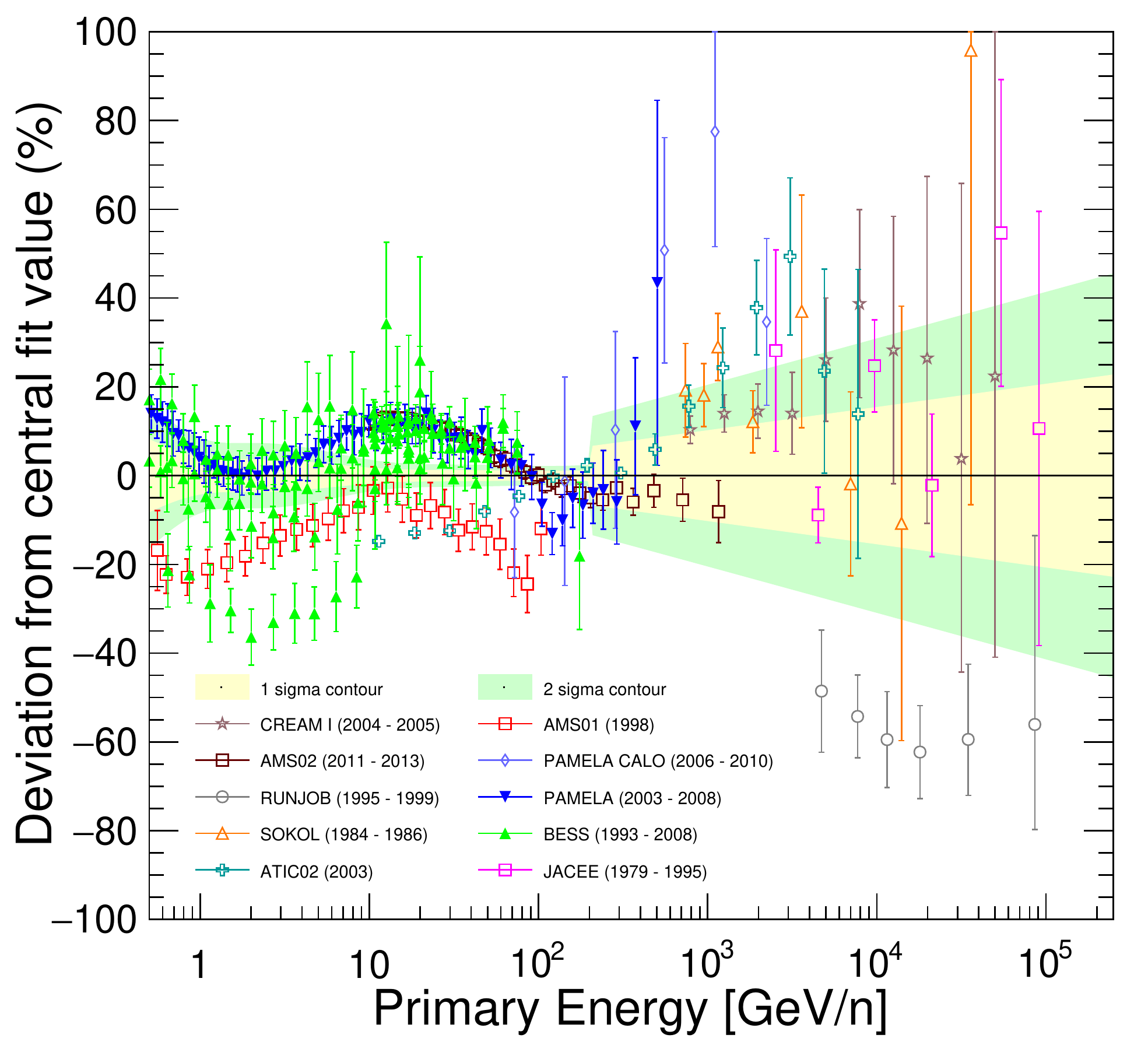}
\caption{Deviation from central fit values of the neutrino flux for (a) proton and (b) helium fluxes as a function of primary energy for the GSHL parametrization. Solid markers indicate spectrometer measurements, empty markers indicate calorimeter measurements. The yellow and green regions indicate the $1$ and $2$~standard deviation contour regions, respectively. The discontinuity above 
an energy of $200$~GeV/n indicates an increase on the uncertainty on the parameter $d$ due to the calorimeter measurements in this energy region.}
\label{fig:devPlot}
\end{figure*}
Combining the acceptability and mutual compatibility methods enables us to estimate a value for $T$. 
Summing all the contributions from the two tolerance methods in Table~\ref{tab:proton}, we obtain $T_{\text{p}} \approx 13$ for proton and $T_{\text{He}} \approx 43$ for helium fluxes. We determine the uncertainties on the GSHL parameters with these values of $T$ by scaling the $1 \sigma$ contour in  $\chi^2$-parameter space. The $\chi^2$/d.o.f. values determined from the fit are $\myvarProtonChi/\myvarProtonDof = \myvarProtonRedChi$ for protons and $\myvarHeliumChi/\myvarHeliumDof = \myvarHeliumRedChi$ for helium. 
Best-fit values and relative uncertainties for the GSHL parameters are shown in Table~\ref{tab:fitResults} and their correlation matrix in Table~\ref{tab:corrMat}.
 
The four parameters are strongly correlated among each other. The parameters $a$ and $d$, which govern the normalization and spectral index, respectively, have a correlation coefficient of $0.98$. Since these two parameters contribute most to the total atmospheric neutrino uncertainties, their correlation cannot be neglected. 

\begin{table}[htbp]
\begin{tabular}{c c c c}
\hline\hline
Parameter & \multicolumn{2}{c}{$\sigma$} & $w$ \\
& ~This work~ & ~Barr \emph{et al.}~ & \\
\hline
$a$ & $3.9 \%$ & $6.7 \%$ & $0.6$ \\
$b$ & $5.6 \%$ & $1.4 \%$ & $4.0$ \\
$c$ & $17.7 \%$ & $10.5 \%$ & $1.7$ \\
$d$ & $0.3 \%$ & $0.4 \%$ & $0.7$ \\
\hline\hline
\end{tabular}
\caption{Relative uncertainty on the GSHL parameters determined in this work and in Ref.~\protect\cite{Barr:2006it}. Their ratio is used to determine a scaling factor $w$ that is applied to the uncertainty functions.}
\label{tab:scaleFactor}
\end{table}

\begin{figure*}[htbp]
	(a)
	\includegraphics[width=0.465\textwidth]{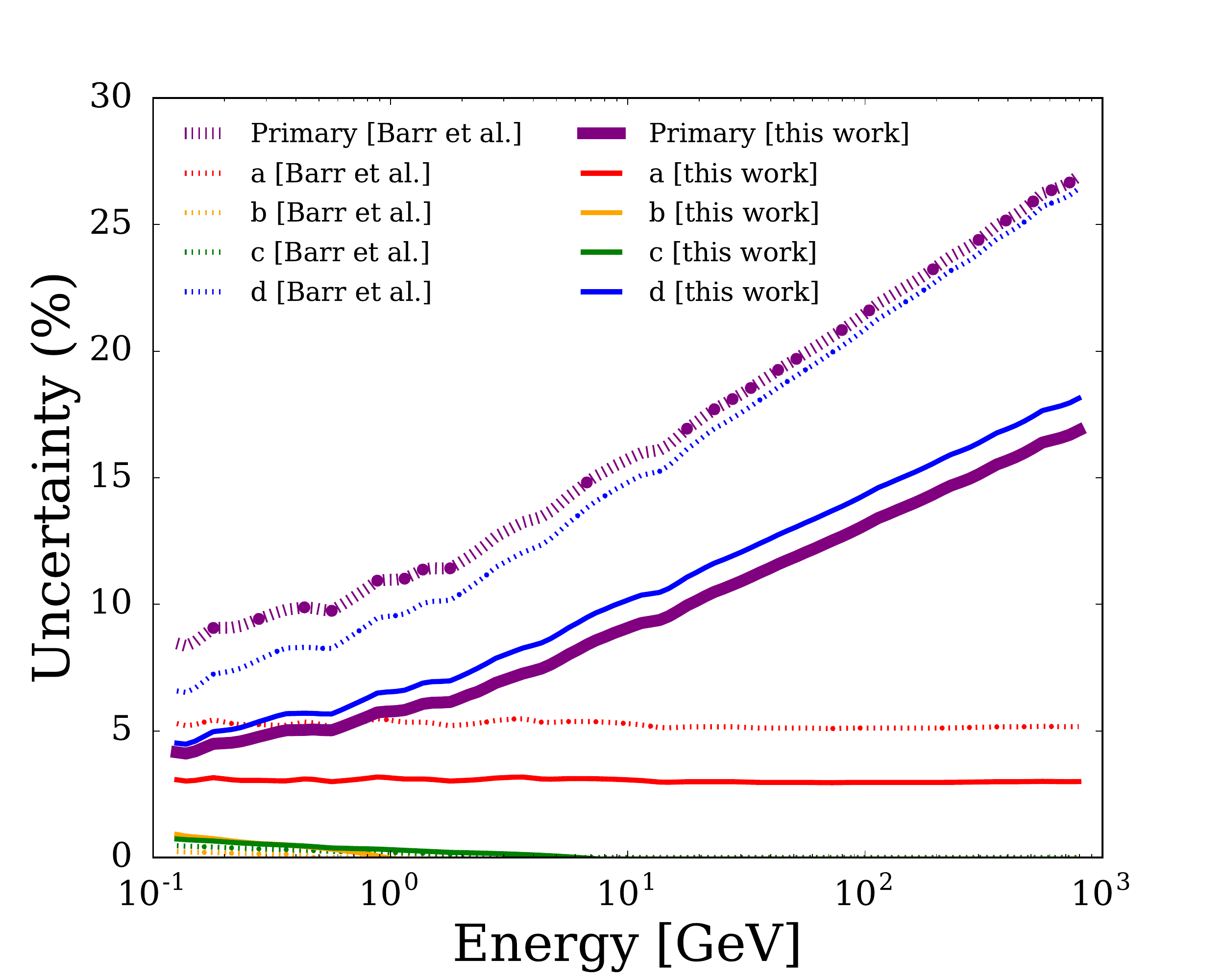}
	(b)
	\includegraphics[width=0.465\textwidth]{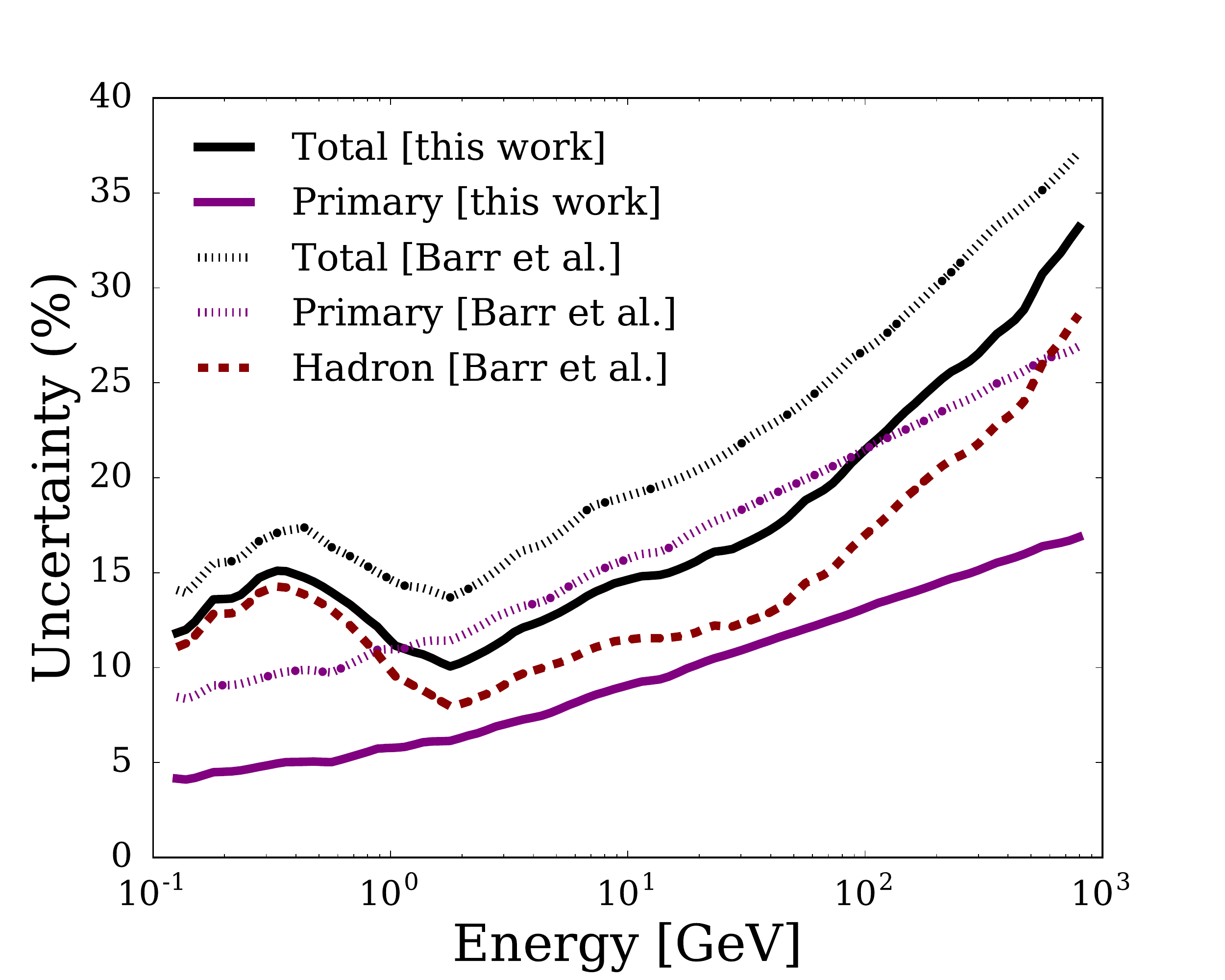}
	\caption{
		Comparison of the (a) individual components and (b) total uncertainties for muon neutrinos between the results of Barr~\emph{et al.}~\protect\cite{Barr:2006it} (dashed) and this analysis (continuous lines).
	}
	\label{fig:newError}
\end{figure*}

\begin{figure*}[htbp]
	(a)
	\includegraphics[width=0.465\textwidth]{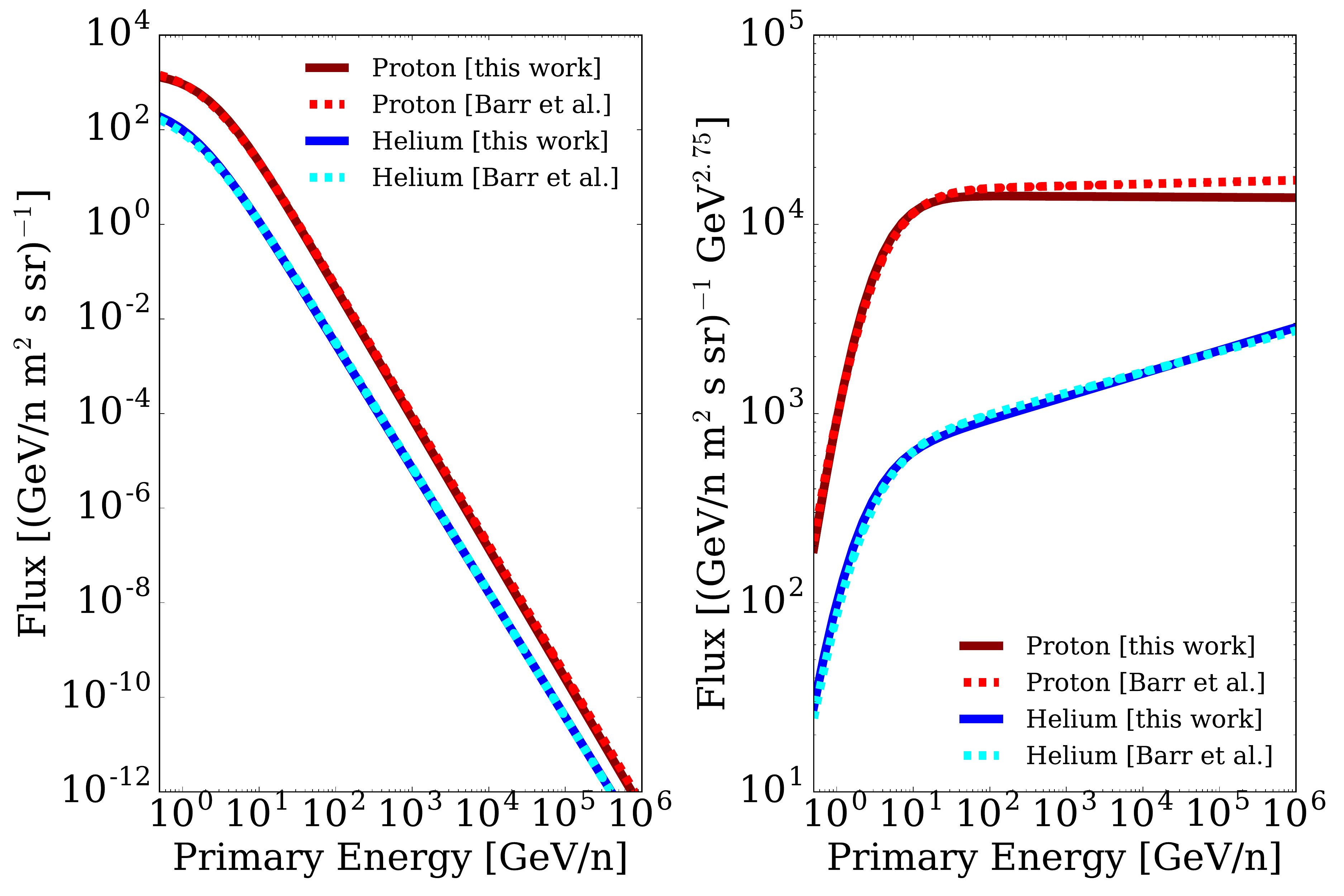}
	(b)
	\includegraphics[width=0.465\textwidth]{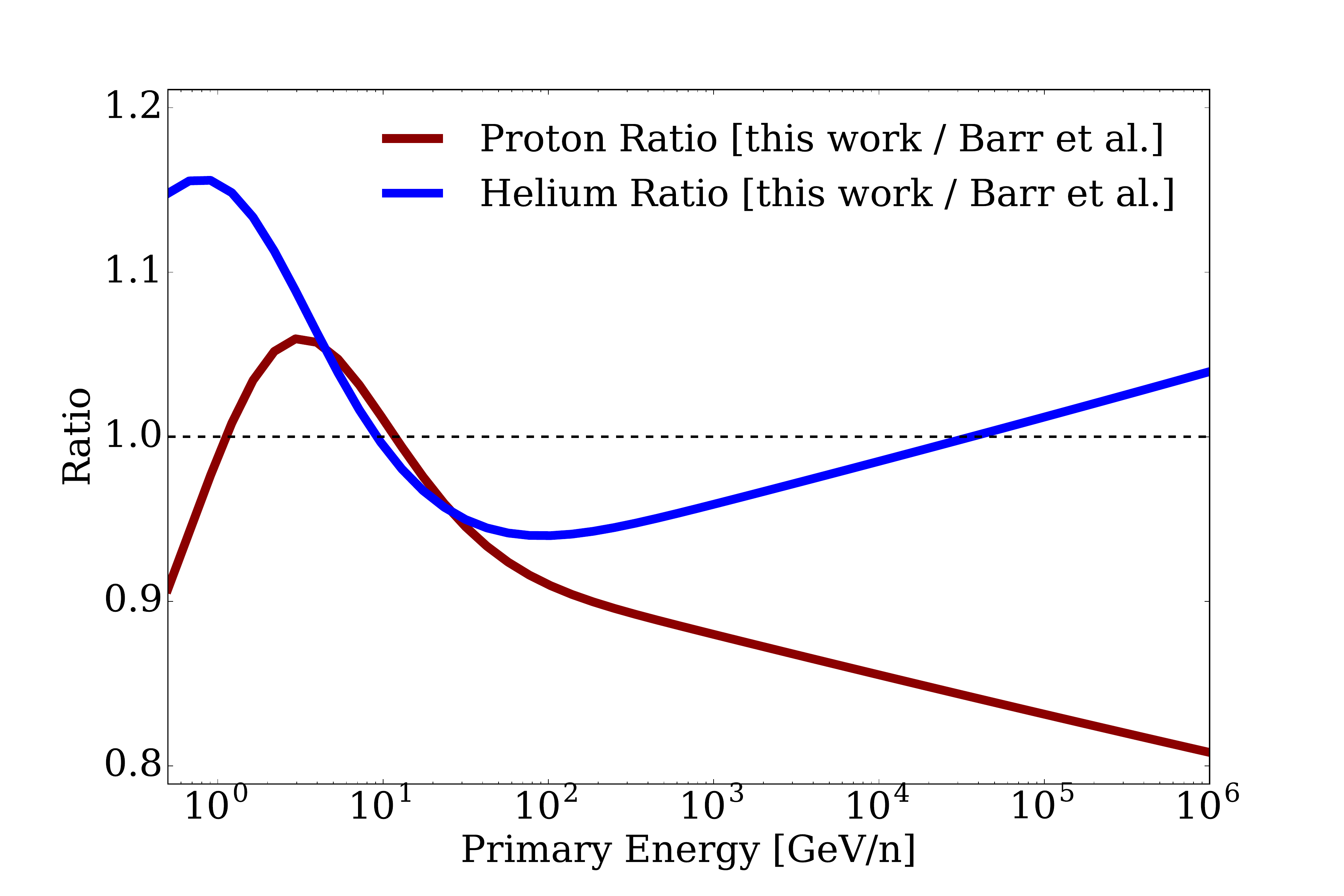}
	\caption{(a) Comparison between the GSHL parameter values from Barr~\emph{et al.}~\cite{Barr:2006it} (dashed line) and our fit (continuous line) for proton and helium primaries; (b) ratio of the two flux parametrisations 
    (this work over Barr~\emph{et al.})}
	\label{fig:porzioBarrRatio}
\end{figure*}

We increase the uncertainty on the parameter $d$ in the energy range above $200$~GeV/n by a factor of 3
for proton and a factor of $2$ for helium fluxes. This maintains consistency with the results of Ref.~\cite{Barr:2006it}, where these uncertainties are inflated to take into account that measurements above $200$~GeV/n were obtained with calorimeter detectors as opposed to spectrometers measuring the energy region below $200$~GeV/n. The inflated uncertainty is carried over into the generation of the uncertainties on the atmospheric neutrino fluxes. We need to adopt the same approach so that
we can perform a rescaling of the uncertainty functions from a direct comparison with the results of Ref.~\cite{Barr:2006it}.

The residuals between data and the fitted GSHL parametrisations are shown in Fig.~\ref{fig:devPlot} against cosmic-ray energy for protons and helium. As Table~\ref{tab:proton} shows, the AMS02 and ATIC data points dominate 
$\chi^2_{\rm{global}}$ and cannot be fitted consistently.
The relative uncertainties on the four parameters obtained from the fit can be compared with the relative uncertainties obtained in Ref.~\cite{Barr:2006it} to determine a scaling factor $w$ that can be applied to the uncertainty functions (see Table~\ref{tab:scaleFactor}). The rescaled uncertainties are shown in Fig.~\ref{fig:newError}.
Figure~\ref{fig:porzioBarrRatio} shows a comparison between the GSHL function using the parameter values extracted by Barr~\emph{et al.}~\cite{Barr:2006it} and our global fit. Our results for protons yield a softer spectrum whereas the helium spectrum is harder.

Barr {\it et al.}~\cite{Barr:2006it} sum the four uncertainty functions in quadrature to obtain a total primary uncertainty for the atmospheric $\nu_{\mu}$ flux. Here, we also take into account the correlations given in Table~\ref{tab:corrMat}. This updated primary uncertainty is summed in quadrature with the hadron production uncertainty determined in Ref.~\cite{Barr:2006it} to obtain a total muon atmospheric neutrino flux uncertainty. The updated total muon-neutrino uncertainties are compared to the previous uncertainties in Fig.~\ref{fig:newError}. 
 We obtain an uncertainty on the muon-neutrino flux related to the primary cosmic rays 
of $\approx (5\text{--}15)\%$, depending on energy, which is about a factor of two smaller than the previously determined uncertainty. 
After adding the hadronic component, the total atmospheric muon-neutrino uncertainty is $\approx 5\%$ lower.

\section{Alternative Parametrisation}
Recent measurements of cosmic-ray spectra for protons and helium nuclei using the AMS-02 detector~\cite{Aguilar:2015ooa,Aguilar:2015ctt} deviate significantly from a single power-law behaviour, exhibiting a harder spectrum above primary energies of $\approx 10^2$~GeV/n. This trend is also observed in the most recent ATIC~\cite{Panov:2011ak} and PAMELA-CALO~\cite{Adriani:2013xva} measurements. 
We therefore propose a modification to the GSHL parametrisation, labeled GSHL+, that includes a shift in the spectral index
using a multi-spectrum parametrisation of the form~\cite{Ter-Antonyan:2014hea}  
\begin{eqnarray}
\label{eq:altgshl}
\lefteqn{\Phi (E_p) = } \nonumber\\
& & \underbrace{a \left[ E_p + b \exp(c \sqrt{E_p})\right] ^{-d}}_{\text{GSHL}} \times \underbrace{\left[  1 + \left( \frac{E_p}{k} \right)^{s} \right]^{\frac{d - e}{s}}}_{\text{Spectral index change}},
\end{eqnarray}
where $a$, $b$, $c$, and $d$ are the parameters used in Eq.~\ref{eq:gshl}, $k$ is the energy per nucleon where the shift in spectral index occurs, $s$ is a sharpness parameter, correlated to the rate of change of the spectral slope, and $e$ is the new spectral index characterising the spectrum at energies above $k$. 

\begin{figure}[htbp]
\includegraphics[width=0.465\textwidth]{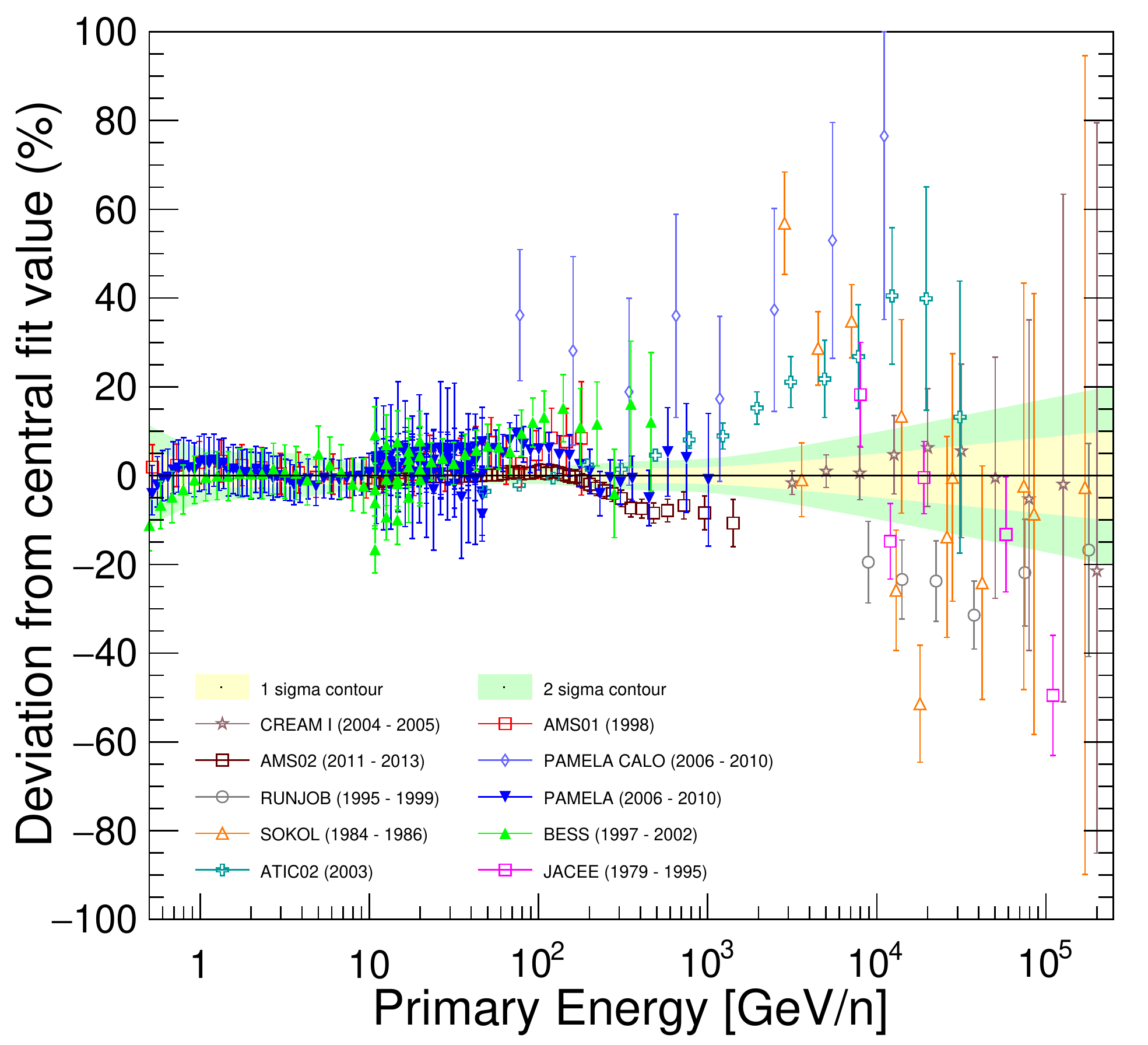}
\caption{Deviation from central fit value for (a) proton and (b) helium fluxes as a function of primary energy for the alternative multi-spectrum parametrization. Solid markers indicate spectrometer measurements, open markers indicate calorimeter measurements. The yellow and green bands represent the $1 \sigma$ and $2\sigma$ contour regions, respectively.}
\label{fig:altDevPlot}
\end{figure}

\begin{figure*}[htbp]
	(a)
	\includegraphics[width=0.465\textwidth]{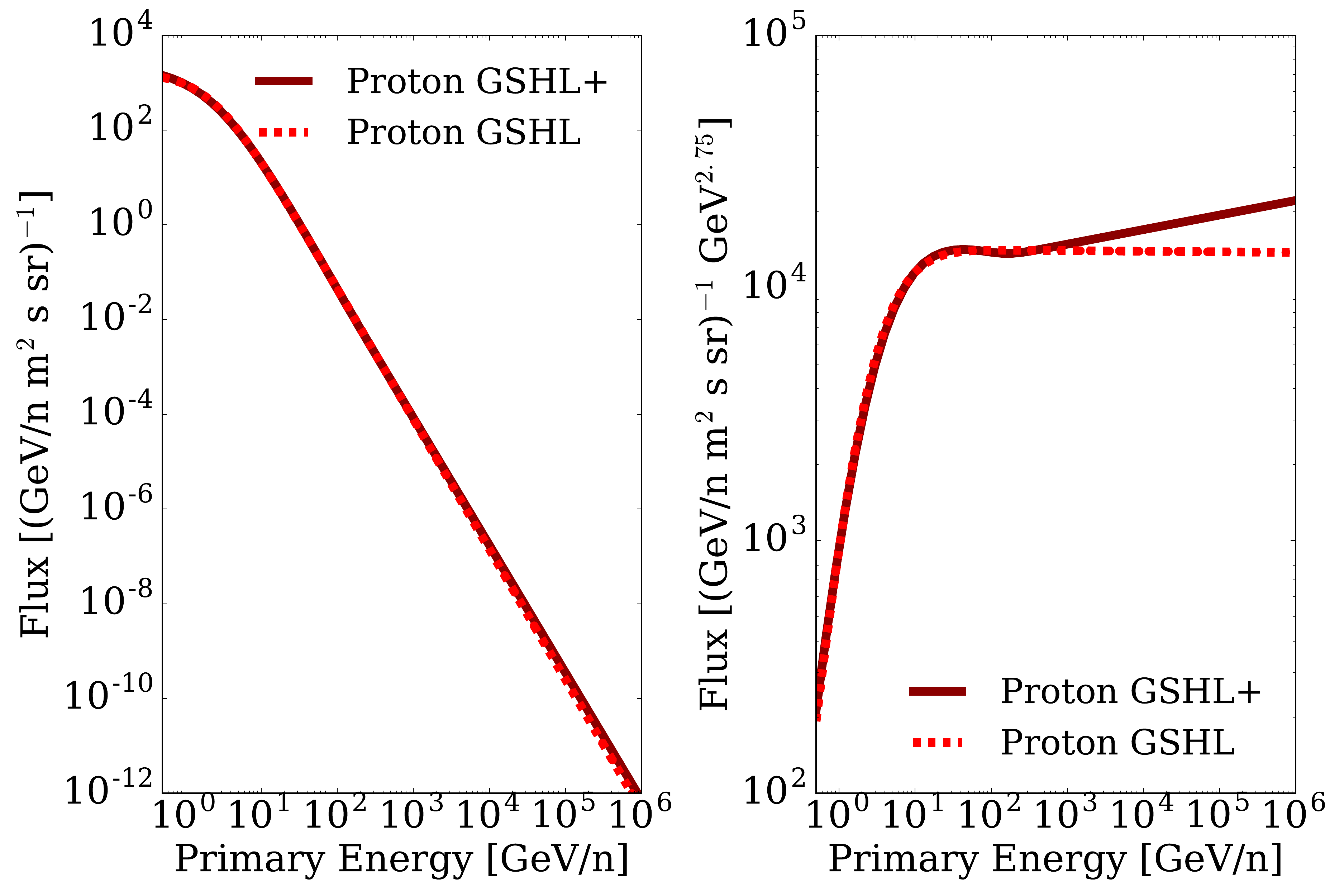}
	(b)
	\includegraphics[width=0.465\textwidth]{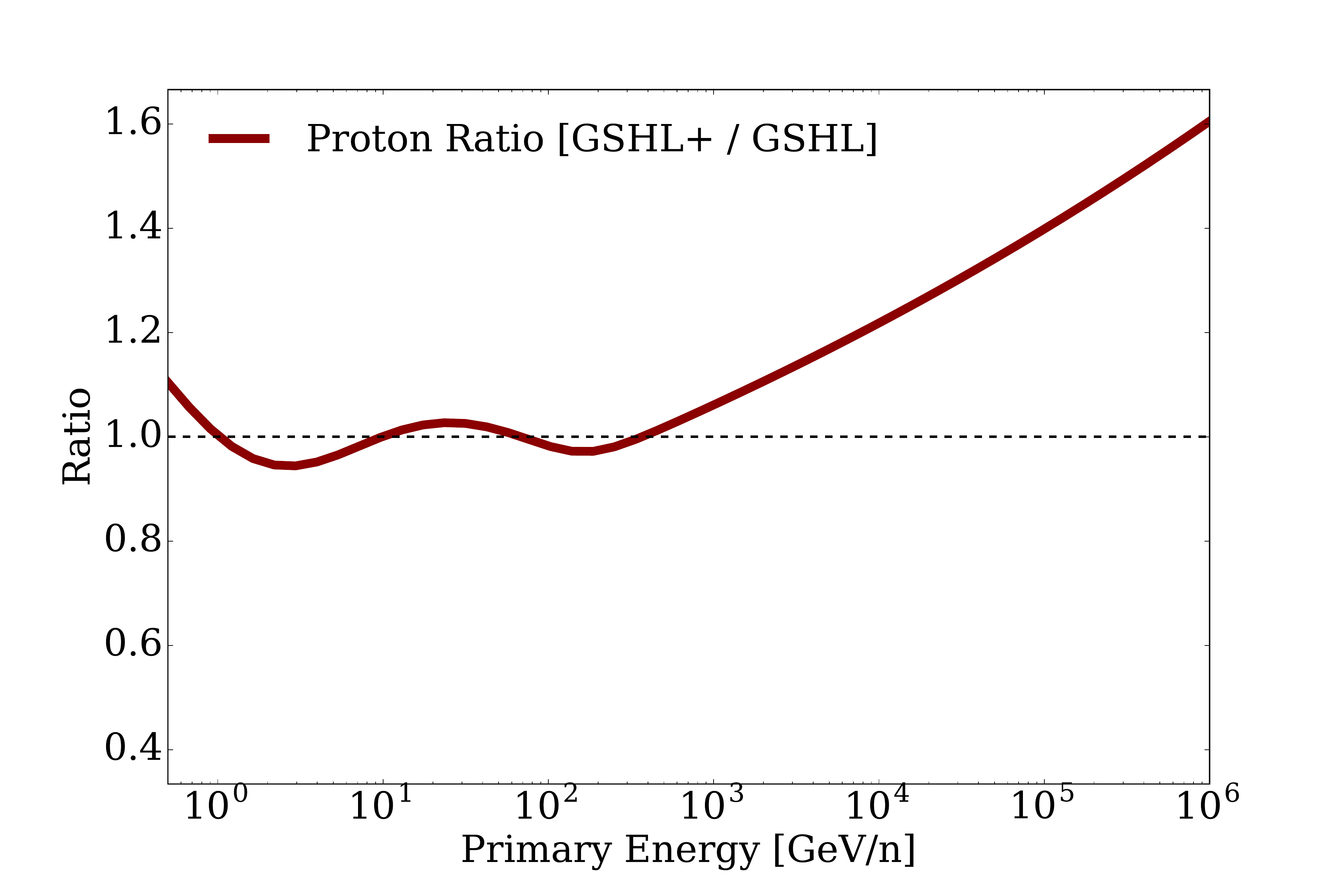}
	\caption{(a) Comparison between the GSHL parametrisation (dashed line) and the alternative multi-spectrum parametrisation (GSHL+, continuous line) for proton; (b) ratio of the GSHL and GSHL+ parametrisations.}
	\label{fig:GSHL+Ratio}
\end{figure*}

The results of the fit are shown in Table~\ref{tab:altFitResults}, and the deviation versus energy in Fig.~\ref{fig:altDevPlot}.
The fit yields $\chi^2/\text{d.o.f.} = \myvarProtonPlusRedChi$ for protons, an improvement compared to $\chi^2/\text{d.o.f.}=\myvarProtonRedChi$ for the GSHL parametrisation.
In Fig.~\ref{fig:GSHL+Ratio} we compare the GSHL function using the parameters of Barr~\emph{et al.}~\cite{Barr:2006it} and our global fit using Eq.~\ref{eq:altgshl}. The results for protons now yield a harder spectrum, which is consistent with the recent observations by the AMS-02, ATIC and PAMELA-CALO Collaborations.  However, fitting the GSHL+ parametrisation for helium returns parameter values that make the extra term disappear (the two spectral indices $d$ and $e$ take approximately the same value) since the helium data is not constraining enough, i.e., there is no benefit to be gained from using the GSHL+ parametrisation.

\begin{table}[htbp]
\begin{tabular}{c c}
\hline\hline
Parameter & Proton \\
\hline
$a$ & $21741 \pm 4707\phantom{0}$ \\
$b$ & $2.29 \pm 0.08$ \\
$c$ & $-0.12 \pm 0.09\phantom{-}$\\
$d$ & $2.85 \pm 0.05$ \\
$k$ & $158.68 \pm 43.70\phantom{0}$ \\
$s$ & $3.56 \pm 3.15$ \\
$e$ & $2.69 \pm 0.02$ \\
\hline\hline
\end{tabular}
\caption{Results from the global fit and uncertainties for the parameters of the alternative parametrisation. Uncertainties are determined using the method from Ref.~\protect\cite{Stump:2001gu,Pumplin:2001ct}. The units are chosen to give the flux $\Phi$ in units of $[(\text{GeV}/\text{n})^{-1}\text{m}^{-2}\text{s}^{-1}\text{sr}^{-1}]$}
	\label{tab:altFitResults}
\end{table}

\begin{table}[htbp]
\begin{tabular}{c | c c c c c c c}
\hline\hline
Parameter & $a$ & $b$ & $c$ & $d$ & $k$ & $s$ & $e$ \\
\hline
$a$ & $\phantom{-}1\phantom{.00}$  & $\phantom{-}0.04$ & $\phantom{-}0.96$  & $\phantom{-}1.00$ & $-0.62$ & $-0.59$ & $\phantom{-}0.00$ \\
$b$ & & $\phantom{-}1\phantom{.00}$ & $-0.22$  & $\phantom{-}0.10$ & $-0.28$ & $-0.30$ & $-0.04$ \\
$c$ & & & $\phantom{-}1\phantom{.00}$ & $\phantom{-}0.93$ & $-0.50$ & $-0.46$ & $\phantom{-}0.02$ \\
$d$ & & & & $\phantom{-}1\phantom{.00}$ & $-0.65$ & $-0.64$ & $-0.00$ \\
$k$ & & & & & $\phantom{-}1\phantom{.00}$ & $\phantom{-}0.27$ & $-0.53$ \\
$s$ & & & & & & $\phantom{-}1\phantom{.00}$ & $\phantom{-}1.00$ \\
$e$ & & & & & & & $\phantom{-}1\phantom{.00}$ \\
\hline\hline
\end{tabular}
\caption{Correlation matrix for the alternative parametrisation.}
\label{tab:altcorrMat}
\end{table}

\section{Discussion}
We show in Sec.~\ref{sec:results} that the GSHL parameters are highly correlated and that a correct treatment of correlations is necessary to obtain a realistic estimate of the total uncertainty. The largest contributions to the total uncertainty from the
parameters $a$ and $d$ exhibit a lower value than the ones presented in Ref.~\cite{Barr:2006it} with different factors. The overall total atmospheric muon-neutrino uncertainty is generally lower than the one previously determined in Ref.~\cite{Barr:2006it}.

The features shown by the most recent high-energy measurements (AMS-02, ATIC, PAMELA-CALO) indicate a harder spectrum above an energy of $\approx 100$ GeV/n than the one predicted by a single power-law spectrum fit. Introducing extra terms for the spectral index shifting results in a lower $\chi^2$ for proton with $\chi^2$ going from \myvarProtonRedChi ~to \myvarProtonPlusRedChi. Due to the action of the extra parameters it is not possible to determine the total flux uncertainty with the rescaling method used for GSHL parametrisation and a full Monte Carlo simulation would be required. However, the results from the fit suggest that a possible estimate of the uncertainty based on the flux generation could benefit from the introduction of the alternative parametrisation. A full treatment of the total atmospheric neutrino uncertainty should also consider updates on the uncertainty from hadron production, which is beyond the scope of this paper.

In summary, the procedure presented in this work employs a more robust method for the determination of the uncertainties associated with the atmospheric neutrino flux, which will facilitate future updates. We also study an alternative parametrisation, which improves the description for energies $>100$~GeV/n. Including recent data sets and taking correlations into account reduces the primary-flux related uncertainty
by about a factor of two, yielding an uncertainty of $(5\text{--}15)\%$. Adding the hadron production uncertainties gives a total uncertainty in the range $(10\text{--}30)\%$. 

\section*{Acknowledgements}
We would like to thank Giles Barr (Oxford) for important discussions and input. We acknowledges support by the Science and Technology Facilities Council (STFC) and the Royal Society.

\bibliography{fluxerrors}
\end{document}